\documentclass[a4paper,11pt]{article}
\usepackage{amscd,amsmath,amssymb,pos}
\usepackage{xcolor,hyperref}

\newcommand{\p}[1]{(\ref{#1})}

\newcommand{\cH}{{\cal H}}

\newcommand{\cN}{{\cal N}}
\newcommand{\cQ}{{\cal Q}}

\newcommand{\cbQ}{\overline{\cal Q}}

\newcommand{\bQ}{{\overline Q}{}}

\newcommand{\bxi}{{\bar\xi}}
\newcommand{\bchi}{{\bar\chi}}
\newcommand{\bpsi}{{\bar\psi}{}}

\newcommand{\und}{\qquad\textrm{and}\qquad}
\renewcommand{\=}{\ =\ }

\newcommand{\be}{\begin{equation}}
\newcommand{\ee}{\end{equation}}
\newcommand{\bea}{\begin{eqnarray}}
\newcommand{\eea}{\end{eqnarray}}

\newcommand{\ba}{\begin{array}} \newcommand{\ea}{\end{array}}

\def\im{{\rm i}}
\def\sfrac#1#2{{\textstyle\frac{#1}{#2}}}

\newcommand{\nn}{\nonumber}
\def\theequation{\arabic{section}.\arabic{equation}}

\begin{document}

\title{New approach to $\cN{=}\,2$ supersymmetric Ruijsenaars--Schneider model}

\author*[a]{Nikolay~Kozyrev}
\author[a]{Sergey~Krivonos}
\author[b]{Olaf~Lechtenfeld}

\affiliation[a]{Bogoliubov  Laboratory of Theoretical Physics, JINR \\ Joliot-Curie 6, 141980 Dubna, Russia}
\affiliation[b]{Institut f\"ur Theoretische Physik and Riemann Center for Geometry and Physics,\\ Leibniz Universit\"at Hannover \\ Appelstrasse 2, 30167 Hannover, Germany}

\emailAdd{nkozyrev@theor.jinr.ru}
\emailAdd{krivonos@theor.jinr.ru}
\emailAdd{olaf.lechtenfeld@itp.uni-hannover.de}

\abstract{We present a very simple form of the supercharges and the Hamiltonian of $\cN{=}\,2$ supersymmetric extension of $n$-particle Ruijsenaars--Schneider models for three cases of the interaction: $1/(x_i-x_j)$, $1/tan(x_i-x_j)$, $1/tanh(x_i-x_j)$. The long ``fermionic tails'' of the supercharges and Hamiltonian rolled up in the simple rational functions depending on fermionic bilinears.}

\FullConference{%
  RDP online workshop "Recent Advances in Mathematical Physics" - Regio2020,\\
  5-6 December 2020\\
  online
}



\maketitle

\setcounter{page}{1}
\setcounter{equation}{0}
\section{Introduction}
The Ruijsenaars-Schneider models \cite{RS1} were introduced by S.N.M.~Ruijsenaars and H.~Schneider who were trying to find a model which could properly describe scattering of relativistic instantons. They studied the model with general pair-wise shift-invariant interactions and found that if the Hamiltonian is a part of $d=2$ Poincar\'e algebra, the interaction is given by the Weierstrass elliptic function. They also noted that the resulting system is integrable and in particular limit can be reduced to the known Calogero-Moser model \cite{Calogero,Calogero1}. The eigenfunctions of this system were found in \cite{Diejen} and are related to the Macdonald polynomials \cite{Macdonald}.

A number of attempts were made to construct a supersymmetric version of this system, all concentrated on $\cN =2$ supersymmetry. In \cite{SRS}, a version of the model was constructed which featured many of its desirable properties, such as integrability, relativistic symmetry, existence of non-relativistic limit which leads to $\cN =2$ supersymmetric Calogero models \cite{susyCal, susyCal1}. This, however, came at the price of very complicated Dirac brackets and conjugation rules for the fermions. Also, of all possible interactions only the trigonometric $1/\sin(x_i-x_j)$ interaction was studied.

In another proposal, A. Galajinsky \cite{ag} used relatively simple supercharges, at most cubic in the fermions, with simple brackets, and limited consideration to a restricted set of possible interactions. Within this approach it was found that $\cN =2$ supersymmetry is possible to achieve for at most three interacting particles. Third construction \cite{KL2} was focused on the same set of functions as the attempt \cite{ag} and relied heavily on modification of brackets of the fermions with themselves and momenta to reproduce the interaction. The conjugation rules of the fermions, unlike \cite{SRS}, were standard. As an unforeseen result, it was found that superalgebra actually closes for any interaction, given by an arbitrary antisymmetric function.

The version of the model we wish to describe in this article involves the fermions with standard Dirac brackets and conjugation rules but the supercharges with specifically prescribed nonlinear dependence on the fermions. As a result, it is possible to close the $\cN =2$, $d=1$ Poincare superalgebra for only a limited set of interactions, all of which belong to the domain of integrability.

This paper is organized as follows. At first, we provide a very brief introduction to the bosonic Ruijsenaars-Schneider model and provide some results of paper \cite{KL2} for later comparison. Second, we introduce the new supercharges that contain a rational function of the fermionic variables and derive the necessary condition that allows to properly close the $\cN =2$, $d=1$ Poincar\'e superalgebra. The details of solution of this equation are given in the Appendix A. Then we relate the obtained supercharges to those found in \cite{KL2}. The details of this calculation are given in the Appendix B.

\setcounter{equation}{0}
\section{General construction}
The Ruijsenaars--Schneider models are integrable many-body systems in one dimension which are described by the equations of motion \cite{RS1}
\be\label{ieq1}
\ddot{x}_i \= 2 \sum^n_{j\neq i} \dot{x}_i \dot{x}_j W(x_i{-}x_j)\ ,
\ee
where the function $W$ is one of the following functions\footnote{
We will not consider the elliptic variant in this paper.}
\be\label{W}
W(x)\ \in\ \bigl\{ 1/x, \ 1/\sin(x), \ 1/\sinh(x), \ 1/\tan(x), \ 1/\tanh(x) \bigr\} \ .
\ee
Equation of motion \p{ieq1} can be reproduced by the Hamiltonian
\be\label{S1}
H \=\sfrac{1}{2} \sum^n_i e^{2\theta_i} \prod^n_{j (\neq i)} f(x_i{-}x_j)\ ,
\ee
called in \cite{RS1} ``light cone Hamiltonian''. Here, the rapidities $\theta_i$ and the coordinates $x_j$ obey standard Poisson brackets,
\be\label{PB0}
\left\{ x_i , \theta_j \right\} =\delta_{ij} \und \left\{x_i,x_j \right\} = \left\{ \theta_i, \theta_j\right\} =0\ .
\ee
The functions $W$ in \p{ieq1} and $f$ in \p{S1} are related (in this order):
\bea\label{fW}
&f(z) \,\in\,\Bigl\{ \frac{1}{z} , \frac{1}{\sinh(z)}, \frac{1}{\sin(z)}, \frac{1}{\tanh(\frac{z}{2})}, \frac{1}{\tan(\frac{z}{2})} \Bigr\}&\;\;\Leftrightarrow\;\; \nn \\
&W(z)\,\in\,\Bigl\{ \frac{1}{z}, \frac{1}{\tanh(z)}, \frac{1}{\tan(z)}, \frac{1}{\sinh(z)}, \frac{1}{\sin(z)} \Bigr\}\ .&
\eea

We prefer the following re-interpretation of the Ruijsenaars--Schneider  systems.
Let us cast the Hamiltonian \p{S1} into a free form by redefining momenta,\footnote{
This form of the Hamiltonians and Poisson brackets is explicitly written in \cite{ag} but may be older.}
\be\label{Hb}
p_i \= e^{\theta_i} \prod^n_{j (\neq i)} \sqrt{f(x_i{-}x_j)}\ \;\; \Rightarrow \;\; H \= \sfrac{1}{2} \sum_{i=1}^n p_i^2\ .
\ee
This redefinition ($\theta_i\to p_i$) clearly changes the Poisson brackets to
\be\label{case1}
\left\{ x_i, p_j\right\} = \delta_{ij} p_j \und \left\{ p_i, p_j\right\}=\left(1-\delta_{ij}\right)p_i p_j W(x_i{-}x_j)\ .
\ee

One may check that the Hamiltonian $H$ \p{Hb} and brackets \p{case1} result in the equations of motion \p{ieq1}. Indeed, from   \p{Hb} and \p{case1} we have
\be\label{eom1a}
\dot{x}_i \ \equiv\  \left\{ x_i, H\right\} \= p_i^2\;\; \Rightarrow \ddot{x}_i \ \equiv\  \left\{ \dot{x}_i, H\right\} \=  2 \sum_{j (\neq i)}^n p_i^2 p_j^2 W(x_i{-}x_j) \= 2  \sum_{j (\neq i)}^n \dot{x}_i \dot{x} _j W(x_{i}{-}x_j),
\ee
as it should be.

The $\cN{=}\,2$ supersymmetric extension of the Ruijsenaars--Schneider models is equivalent to the existence of supercharges
 $\cQ$ and $ \cbQ$ forming an $\cN{=}\,2$ Poincar\'{e} superalgebra
\be\label{N2sa}
\left\{ \cQ, \cbQ \right\} = - 2 \im \cH \und \left\{ \cQ, \cQ \right\}=\left\{ \cbQ, \cbQ \right\}=0
\ee
together with the Hamiltonian $\cH$ whose bosonic sector coincides with the Hamiltonian $H$ \p{Hb}.

To construct such supercharges the  $2\,n$ phase-space variables $x_i$ and $p_j$,
obeying the brackets \p{case1}, have been extended in \cite{KL2}  by  $2\,n$  fermions $\xi_i$ and $\bxi_{j}=\left(\xi_j\right)^\dagger$ , subject to the standard brackets
\be\label{PBxi}
\left\{\xi_i, \xi_j\right\} =\left\{\bxi_i, \bxi_j\right\}=0\ ,\quad
\left\{\xi_i, \bxi_j\right\} =-\im\, \delta_{ij} \und
\left\{p_i, \xi_j\right\}= \left\{p_i, \bxi_j\right\}=0\ .
\ee

The supercharges $\cQ, \cbQ$ and the Hamiltonian $H$ obeying the $\cN{=}2$ superalgebra \p{N2sa} were constructed in \cite{KL2} as
\bea\label{QQH1}
\cQ &=& \sum_{i}^n p_i \exp\Bigl\{- \frac{\im}{2}\, \sum_j^n \xi_j \bxi_j W(x_i{-}x_j)\Bigr\}\ \xi_i \ ,\;\;
\cbQ \= \sum_{i}^n p_i  \exp\Bigl\{ \frac{\im}{2}\, \sum_j^n \xi_j \bxi_j W(x_i{-}x_j)\Bigr\} \bxi_i\ ,
\nn \\
\cH&=& \sfrac{1}{2} \sum_{i=1}^n p^2_i \ +\ \im \sum_{i,j}^n p_i p_j e^{-\frac{i}{2} \sum_k \xi_k \bxi_k \left(W(x_i{-}x_k)-W(x_j{-}x_k)\right)}\xi_i \bxi_j W(x_i{-}x_j) \nn \\
&&
\ +\ \sfrac{1}{2} \sum_{i,j}^n p^2_i \xi_i \bxi_i \xi_j\bxi_j W'(x_i{-}x_j).
\eea

One should stress that the supercharges and the Hamiltonian $H$ \p{QQH1} obey superalgebra \p{N2sa} for arbitrary odd potential $W(x)$, without restriction \p{W} to have an integrable bosonic Hamiltonian.

\setcounter{equation}{0}
\section{Alternative $\cN{=}2$ supersymmetric  Ruijsenaars-Schneider model}
Trying to devise alternative ways to construct $\cN{=}2$ supersymmetric Ruijsenaars-Schneider model, one can consider the variables with following simple Dirac brackets \p{case1}, \p{PBxi}
\be\label{brackets}
\{ x_i, p_j \} =\delta_{ij}p_j, \;\; \{ p_i, p_j  \} = W(x_i-x_j)p_i p_j, \;\; \{ \xi_i, \bxi_j \} =- \im \delta_{ij}
\ee
(omitted brackets are equal to zero) and focus on modifying the structure of the supercharges. For example, one can consider supercharges written as power series in the matrix
\bea\label{Pi}
\Pi_{ij} = \frac{\im}{2}\, W(x_i-x_j)\big(\xi_i \bxi_j -\xi_j \bxi_i\big), \qquad \left( \Pi_{ij} \right)^\dagger =\Pi_{ij},\quad \Pi_{ij}=\Pi_{ji}, \nn \\
Q = \sum_{i,j}p_i \xi_j \Big(\delta_{ij}+a_1 \Pi_{ij} +a_2 \sum_k \Pi_{ik}\Pi_{kj} +a_3 \sum_{k,m} \Pi_{ik}\Pi_{km}\Pi_{mj} +\ldots \Big).
\eea
To check whether such an assumption is useful, one can try to satisfy $\cN{=}2$, $d=1$ super Poincar\'e condition $\big\{ Q,Q  \big\}=0$ for systems with small number (2,3 or 4) of particles and simplest interaction $W(x)=1/x$. Then one quickly arrives at the conclusion that, indeed, $\big\{ Q,Q  \big\}=0$ can be achieved if one sets $a_1=a_2 = \ldots =1$. Such supercharges can be rather compactly written as
\be\label{QQb1}
Q= \sum_{i,j=1}^n p_i \left(\frac{1}{1-\Pi}\right)_{ij} \xi_j ,\;\;
\bQ= \sum_{i,j=1}^n p_i  \left(\frac{1}{1-\Pi}\right)_{ij} \bxi_j\ .
\ee
Intriguing feature of this representation is that, unlike the previously considered supercharges \p{QQH1}, they do not form $\cN{=}2$ Poincar\'e superalgebra \p{N2sa} for an arbitrary interaction $W(x)$. Considering only integrable interactions, one finds that $\big\{ Q,Q  \big\}=0$ is satisfied also for $W(x)=1/\tan(x)$ and $W(x)=1/\tanh(x)$, but not for $W(x)=1/\sin(x)$ and $W(x)=1/\sinh(x)$. Such a strong condition is not typical for $\cN{=}2$ supersymmetric mechanics.

To find out the exact condition the $W(x)$ function should satisfy, one should study the expansion of $\big\{ Q,Q  \big\}$ bracket as a power series in the fermions. The terms of second and fourth power in the fermions cancel automatically due to choice of coefficients in the power series. However, the 6th power cancels only if a cubic equation is satisfied:
\bea\label{cubiceq}
E_{[ik](jl)} =W_{ij}W_{ik}W_{il}-W_{ij}W_{il}W_{jk}+W_{ik}W_{il}W_{jk}-W_{il}W_{jk}W_{jl}-W_{ij}W_{ik}W_{kl}+\nn \\+W_{ij}W_{il}W_{kl}
+W_{ij}W_{jk}W_{kl}-W_{ik}W_{jk}W_{kl}+W_{il}W_{jk}W_{kl}-W_{ij}W_{jl}W_{kl}=0.\nn
\eea
Here, to make notation shorter, we introduced $W_{ij}=W(x_i-x_j)$. Careful analysis shows that this equation is actually equivalent to the quadratic one. Indeed, solving together $E_{[ik](jl)}=0$ and $E_{[jk](il)}=0$ with respect to $W_{ik}$ and $W_{il}$, one finds that
\bea\label{cubiceqsol1}
&W_{ik} = \frac{W_{ij}W_{jk}+W_{jk}W_{jl}-W_{jk}W_{kl}+W_{jl}W_{kl}}{W_{ij}+W_{jk}}, \;\;
 W_{il} = \frac{W_{ij}W_{jl}+W_{jk}W_{jl}-W_{jk}W_{kl}+W_{jl}W_{kl}}{W_{ij}+W_{jl}},& \nn \\
&\mbox{or}\;\; W_{ik}W_{ij}+W_{ik}W_{jk}+W_{ji}W_{jk} = W_{jk}W_{jl}+W_{kj}W_{kl}+W_{jl}W_{kl}.&
\eea
One can observe that each side of last equation is symmetric with respect to permutation of arguments, and the left hand side depends on $x_i$, $x_j$, $x_k$ only, while the right hand side does not depend on $x_i$ but depends on $x_l$. This could happen if and only if both sides are equal to the same constant. Thus we find the main condition that function $W(x)$ should satisfy
\be\label{maineq}
W(x_i-x_j)W(x_i-x_k)+W(x_j-x_i)W(x_j-x_k)+W(x_k-x_i)W(x_k-x_j)=c=\mbox{const}.
\ee
One can note that the only significant values of $c$ are $c=-1,0,1$, as the solutions with another values of $c$ can be obtained by rescaling of $W$. The only nontrivial solutions to equation \p{maineq} are\footnote{We put a sketch of the proof in the Appendix A}
\be\label{solm}
W(x) = \left\{ \begin{array}{l}
 \frac{1}{x}, \qquad c = 0\\ [4pt]
 \frac{1}{\tan x}, \qquad c = -1 \\ [4pt]
 \frac{1}{\tanh x},  \qquad c = 1
\end{array} \right. .
\ee
Let us again stress that it is a rather nontrivial fact that $\cN{=}2$ supersymmetry with the fixed form of the supercharges \p{QQb1} select just three cases with the integrable bosonic Hamiltonians. Usually, $\cN{=}2$ supersymmetry is too weak to select integrable systems. For example, the supercharges constructed in \cite{KL2} we started with \p{QQH1} form $\cN{=}2$ superalgebra \p{N2sa} for arbitrary potential $W(x)$.

To prove that the supercharges satisfy the $\cN{=}2$ superalgebra condition in all orders in the fermions, one definitely should take into account that the only acceptable functions $W(x)$ are listed \p{solm}. The simplest way to perform this is to connect the present construction to the previous work \cite{KL2}. Indeed, performing resummation of power series, one could prove that the proposed supercharges \p{QQb1} coincide with the ``exponential'' ones \p{QQH1} for $W(x)=1/x$, as
\bea\label{Qrat}
&&Q_{rat} = \sum_{i,j=1} p_i \left(\frac{1}{1-\Pi}\right)_{ij} \xi_j = \sum_i p_i\exp\Bigl\{- \frac{\im}{2}\, \sum_j^n \frac{\xi_j \bxi_j}{x_i{-}x_j}\Bigr\}\ \xi_i\, =\cQ_{rat} ,\nn \\
&&\bQ_{rat}=\sum_{i,j=1} p_i  \left(\frac{1}{1-\Pi}\right)_{ij} \bxi_j =\sum_i p_i \exp\Bigl\{ \frac{\im}{2}\, \sum_j^n \frac{\xi_j \bxi_j}{x_i{-}x_j}\Bigr\} \bxi_i =\cbQ_{rat},\nn \\
&&\Pi_{ij} = \frac{\im}{2}\frac{\xi_i\bxi_j -\xi_j \bxi_i}{x_i-x_j}.
\eea
The explicit proof of this relation is given in the first half of the Appendix B.

In the trigonometric/hyperbolic case the relation is more complicated, and the supercharges differ by multiplication by functions of $J= \sum_k \xi_k\bxi_k$, which, however, do not spoil the superalgebra. The explicit relations between the ``trigonometric'' supercharges read
\bea\label{Qtrig}
&& Q_{tan} = \sum_{i,j} p_i \left( \frac{1}{1-\Pi}  \right)_{ij}\xi_j  =\frac{1}{\cosh\left( \frac{J}{2} \right)} \sum_{i}^n p_i \exp\Bigl\{- \frac{\im}{2}\, \sum_j^n \frac{\xi_j \bxi_j}{\tan(x_i{-}x_j)} \Bigr\}\ \xi_i = \frac{1}{\cosh\left( \frac{J}{2} \right)}\cQ_{tan}, \nn \\
&&\bQ_{tan} = \sum_{i, j} p_i \left( \frac{1}{1-\Pi}  \right)_{ij}\bxi_j  = \frac{1}{\cosh\left( \frac{J}{2} \right)} \sum_{i}^n p_i \exp\Bigl\{ \frac{\im}{2}\, \sum_j^n \frac{\xi_j \bxi_j}{\tan(x_i{-}x_j)} \Bigr\}\ \bxi_i = \frac{1}{\cosh\left( \frac{J}{2} \right)}\cbQ_{tan}, \nn \\
&&\Pi_{ij} = \frac{\im}{2}\, \frac{ \xi_i \bxi_j - \xi_j \bxi_i}{\tan(x_i-x_j)}.
\eea
The relations in the hyperbolic case are almost similar:
\bea\label{Qhyp}
&& Q_{tanh} = \sum_{i,j} p_i \left( \frac{1}{1-\Pi}  \right)_{ij}\xi_j  =\frac{1}{\cos\left( \frac{J}{2} \right)} \sum_{i}^n p_i \exp\Bigl\{- \frac{\im}{2}\, \sum_j^n \frac{\xi_j \bxi_j}{\tanh(x_i{-}x_j)} \Bigr\}\ \xi_i = \frac{1}{\cos\left( \frac{J}{2} \right)}\cQ_{tanh}, \nn \\
&&\bQ_{tanh} = \sum_{i, j} p_i \left( \frac{1}{1-\Pi}  \right)_{ij}\bxi_j =  \frac{1}{\cos\left( \frac{J}{2} \right)} \sum_{i}^n p_i \exp\Bigl\{ \frac{\im}{2}\, \sum_j^n \frac{\xi_j \bxi_j}{\tanh(x_i{-}x_j)} \Bigr\}\ \bxi_i = \frac{1}{\cos\left( \frac{J}{2} \right)}\cbQ_{tanh}, \nn\\
&&\Pi_{ij} = \frac{\im}{2}\, \frac{ \xi_i \bxi_j - \xi_j \bxi_i}{\tanh(x_i-x_j)}.
\eea
The detailed proof of these relations is given in the Appendix B. However, one can give a simpler argument why the $J$-dependent factors should appear. Let us (approximately) calculate the brackets between the generalized fermions in the ``exponential'' basis \cite{KL2}
\be\label{psi21}
\cQ =\sum_i p_i\psi_i, \;\; \cbQ = \sum_i p_i \bpsi_i \;\; \Rightarrow
\left\{ \begin{array}{l}
\psi_i = \exp\Bigl\{- \frac{\im}{2}\, \sum_j^n \frac{\xi_j \bxi_j}{\tan(x_i{-}x_j)}\Bigr\}\ \xi_i \\
\bpsi_i = \exp\Bigl\{ \frac{\im}{2}\, \sum_j^n \frac{\xi_j \bxi_j}{\tan(x_i{-}x_j)}\Bigr\}\ \bxi_i
\end{array} \right. \; , \quad
\left\{ \psi_i, \bpsi_i\right\} = -\im\ .
\ee
At the same time, one can define the generalized fermions in the trigonometric case \p{Qtrig} $Q_{tan}=\sum_i p_i\chi_i$, $\bQ_{tan} = \sum_i p_i \bchi_i$ and approximately calculate their bracket
\bea\label{rel1}
\chi_i =\sum_{j}\left( \frac{1}{1-\Pi}  \right)_{ij}\xi_j, \;\; \bchi_i = \sum_{j}\left( \frac{1}{1-\Pi}  \right)_{ij}\bxi_j, \;\; \Pi_{ij} = \frac{\im}{2}\, \frac{ \xi_i \bxi_j - \xi_j \bxi_i}{\tan(x_i-x_j)} \;\; \Rightarrow \nn \\
\big\{ \chi_i, \bchi_i\big\} = -\im \frac{1}{\cosh^2\big(J/2\big)}+\im \frac{\sinh\big(J/2 \big)}{\cosh^3 \big( J/2\big)}\xi_i\bxi_i.
\eea
Taking into account that $\xi_i\bxi_i = \psi_i\bpsi_i$, one can observe that
\be\label{chibchiredef}
\big\{\cosh(J/2)\chi_i,\cosh(J/2) \bchi_i \big\}=-\im
\ee
and thus one can identify $\chi_i = \psi_i/\cosh(J/2)$, just as in \p{Qtrig}.

Let us also mention that the supercharges \p{QQH1} can be rescaled by an arbitrary function of $J$ and still form $\cN{=}2$, $d{=}1$ Poincar\'e superalgebra. Indeed, as a consequence of brackets \p{PBxi}, $\big\{ \xi_i\bxi_i, \xi_j\bxi_j \big\} =0$ and, therefore, $\big\{f(J),\cQ\big\}=-\im f^\prime(J)\cQ$ and $\big\{ f(J)\cQ, f(J)\cQ \big\} \sim f f^\prime \,\cQ^2=0$.

\section{Conclusions}
In this paper, we constructed a simple $\cN=2$ supersymmetric generalization of the Ruijsenaars-Schneider model. The discussed model is closely related to one constructed in \cite{KL2} but assumes standard brackets (and conjugation rules) of the fermions from the beginning and relies on the specific structure of supercharges to produce the proper interaction terms. Unlike other proposed supersymmetric versions of the Ruijsenaars-Schneider model, which either assumed the specific interaction from the beginning \cite{SRS,ag} or were valid for any interaction given by an odd function \cite{KL2}, our model naturally produces constraints on possible interactions. All three selected interaction functions
\be
W(z)\,\in\,\Bigl\{ \frac{1}{z}, \frac{1}{\tanh(z)}, \frac{1}{\tan(z)} \Bigr\} \nn
\ee
correspond to the models with integrable bosonic core, thus naturally raising the questions whether the supersymmetric model is integrable and how the assumed structure of supercharges is related to integrability. Unfortunately, other integrable interactions $1/\sin(z)$ and $1/\sinh(z)$, as well as more general interaction related to the Weierstrass elliptic function,  fall out of the scope of the present paper. While the present scheme does not seem to be directly generalizable to higher supersymmetries, another question for further study is construction of $\cN=4$ extensions of the model, possibly using ideas of papers \cite{KLS} and \cite{KLPS}.

\acknowledgments

This work was supported by the RFBR-DFG grant No 20-52-12003.

\def\theequation{A.\arabic{equation}}
\setcounter{equation}{0}
\appendix
\section{Solution to the main equation}
To prove that the equation \p{maineq}
\be\label{maineqcopy}
W(x_i-x_j)W(x_i-x_k)-W(x_i-x_j)W(x_j-x_k)+W(x_i-x_k)W(x_j-x_k)=c=\mbox{const}
\ee
has only solutions $W(x)\sim 1/x, 1/\tan x, 1/\tanh x$, depending on value of $c$,
one should note that \p{maineqcopy} is a relation on the function $W$ that should be identically satisfied for any values of three variables $x_i$, $x_j$, $x_k$ within the domain of acceptability ($x_i\neq x_j \neq x_k$). Thus one could let $x_k=0$ to find
\be\label{maineqsol1}
W(x_i-x_j)W(x_i)-W(x_i-x_j)W(x_j)+ W(x_i)W(x_j)=c\; \Rightarrow W(x_i -x_j) = \frac{c - W(x_i)W(x_j)}{W(x_i)-W(x_j)}.
\ee
Substituting this into equation \p{maineqcopy}, one can see that it is satisfied identically. Then, introducing $\varphi(x)=1/W(x)$ and replacing $x_j \rightarrow -x_j$, it becomes evident that \p{maineqsol1} is the law satisfied by the tangent/hyperbolic tangent/linear function of sum of two arguments, depending on value of $c$:
\bea\label{maineqsol2}
\varphi(x_i +x_j) = \frac{\varphi(x_i)+\varphi(x_j)}{1+c\, \varphi(x_i)\varphi(x_j)} \;\; \Rightarrow \;\;
\varphi(x) = \left\{ \begin{array}{l}
  x, \qquad c = 0\\
 \tan x, \qquad c = -1 \\
 \tanh x,  \qquad c = 1
\end{array} \right. .
\eea
There are no more smooth solutions, as one can find that, as a consequence of \p{maineqsol2}, the function $\varphi$ satisfies a differential equation which is enough to constrain the functional form of $\varphi$. Indeed, for general differentiable function
\be\label{maineqsol3}
\varphi(x+\epsilon) = \varphi(x)+\epsilon \, \varphi^\prime(x)+O(x^2).
\ee
At the same time,  \p{maineqsol2} implies
\be\label{maineqsol4}
\varphi(x+\epsilon) = \frac{\varphi(x)+\varphi(\epsilon)}{1+c\,\varphi(x)\varphi(\epsilon)}.
\ee
Treating $\epsilon$ as an infinitesimal parameter, one notes that $\varphi(\epsilon) = \varphi(0)+\epsilon\varphi^\prime(0) + O(\epsilon^2) =a \epsilon + O(\epsilon^2)$. Here, $\varphi(0)=0$, as $\varphi(x)$ is odd, and $a=\varphi^\prime (0)$ is some constant.  Therefore,
\bea\label{maineqsol5}
\varphi(x+\epsilon)-\varphi(x) = \frac{\varphi(x)+\varphi(\epsilon)}{1+c\,\varphi(x)\varphi(\epsilon)}-\varphi(x)  =a\epsilon\big(1-c \,\varphi^2(x)\big) + O(\epsilon^2).
\eea
Thus, as a consequence of \p{maineqsol2},  $\varphi(x)$ satisfies a differential equation with easily obtained solutions
\be\label{maineqsol6}
\varphi^\prime(x) =a\big(1-c \,\varphi^2(x)\big)\; \Rightarrow \; \left\{\begin{array}{ll} c=0\; \Rightarrow \varphi(x)=a\,x+C_0 \\ c=-1\; \Rightarrow \varphi(x)=\tan(a\,x+C_{-1}) \\ c=1\; \Rightarrow \varphi(x)=\tanh(a\,x+C_1) \end{array}\right.
\ee
As the functions should be odd, $C_0$, $C_{-1}$, $C_1$ should be set to zero.  The constant $a$ is unessential as \p{maineqcopy} holds even under arbitrary change of variables $x_i\rightarrow y_i(x_j)$.

\def\theequation{B.\arabic{equation}}
\setcounter{equation}{0}
\section{Relation with the exponential form of supercharges}
To prove the formulas \p{Qrat}, \p{Qtrig}, \p{Qhyp}, we should at first study the general properties of the power series \p{Pi}, \p{QQb1}.

At first, one should note that $\xi_i \big( \Pi^\alpha \big)_{ij}\xi_j =0$. This is obvious for $\alpha=1$ and can be proven for any power $\alpha$ by induction:
\be\label{xipixi}
\xi_i \big( \Pi^\alpha \big)_{ij}\xi_j =\frac{\im}{2} \xi_i \bxi_i \sum_k  W(x_i-x_k)\big( \Pi^{\alpha-1} \big)_{kj}\xi_k\xi_j.
\ee
Therefore,
\be\label{xipixi2}
\sum_j \big( \Pi^{\alpha} \big)_{ij}\xi_j =  \frac{\im}{2}\sum_{j,k} W(x_i-x_k)\big( \xi_i \bxi_k -\xi_k \bxi_i \big)\big( \Pi^{\alpha-1} \big)_{kj}\xi_j =- \xi_i \frac{\im}{2 }\sum_{j,k} W(x_i-x_k)\big(\Pi^{\alpha-1} \big)_{kj} \xi_j \bxi_k
\ee
and one can present $\sum_j(1-\Pi)^{-1}_{ij}\xi_j = \xi_i+\lambda_i\xi_i$ for any $W$.
Then one can derive equation the function $\lambda_i$ should satisfy
\be\label{lambdaeq}
\xi_i = \sum_{j,k}(1-\Pi)_{ij}\left( \frac{1}{1- \Pi}  \right)_{jk}\xi_k \; \Rightarrow \;\xi_i\big(\lambda_i + \frac{\im}{2} \sum_j W(x_i-x_j)\xi_j\bxi_j + \frac{\im}{2} \sum_j W(x_i-x_j)\xi_j\bxi_j \lambda_j\big) =0.
\ee
In more clear notation, it could written as (we assume that $\xi_i$ can be factorized out)
\be\label{lambdaeq1}
\sum_j \big( \delta_{ij} - Z_{ij} \big) \lambda_j = T_i, \;\; Z_{ij} =-\frac{\im}{2}W(x_i-x_j)\xi_j \bxi_j, \;\; T_i = \sum_j Z_{ij} \; \Rightarrow \lambda_i =\sum_{\alpha=0}^{\infty}\sum_j \big( Z^{\alpha} \big)_{ij}T_j.
\ee
These relations were, so far, valid for general $W(x_i-x_j)$. If $W(x_i-x_j)$ is a solution to \p{maineqcopy}, it can be shown that $Z_{ij}$ satisfies the relation
\bea\label{Zeq}
\sum_j Z_{ij}Z_{jk} = \left( -\frac{\im}{2} \right)^2 \sum_j W(x_i-x_j)W(x_j-x_k)\xi_j\bxi_j \, \xi_k\bxi_k = \nn \\
=  \left( -\frac{\im}{2} \right)^2 \sum_j \big( W(x_i-x_j)W(x_i-x_k) - W(x_i-x_k)W(x_k-x_j) -c  \big)\xi_j\bxi_j \, \xi_k\bxi_k \; \Rightarrow \nn \\
\sum_j Z_{ij}Z_{jk} = \big( T_i -T_k \big)Z_{ik} + \frac{1}{4}c\,\xi_k \bxi_k \, J, \;\; J=\sum_m \xi_m \bxi_m.
\eea
This relation should be used to calculate $\sum_j \big( Z^\alpha \big)_{ij}T_j$ to find the solution of \p{lambdaeq1}.

\subsection*{Rational $W$}
In the simpler rational case $c=0$, one can observe that $\sum_j\big(Z^\alpha\big)_{ij}T_j$ for $\alpha=1,2$ reduces to $\big(T_i \big)^2$ and $\big(T_i \big)^3$:
\bea\label{ZTrels}
\sum_j Z_{ij}T_j = \sum_{j,k} Z_{ij}Z_{jk} = \big(T_i\big)^2 - \sum_k Z_{ik}T_k \; \Rightarrow \; \sum_j Z_{ij}T_j = \frac{1}{2} \big(T_i\big)^2, \nn \\
\sum_j \big(Z^2\big)_{ij}T_j = \frac{1}{2}\sum_j Z_{ij}\big(T_{j}\big)^2 = T_{i} \sum_k Z_{ik}T_k -\sum_k Z_{ik} \big(T_{k}\big)^2 \; \Rightarrow \; \nn \\ \sum_j Z_{ij}\big(T_{j}\big)^2 = \frac{1}{3} \big(T_{i}\big)^3, \; \sum_j \big(Z^2\big)_{ij}T_j = \frac{1}{6} \big(T_{i}\big)^3,
\eea
with coefficients coinciding with those in the power expansion of $e^{T_i}$.
To find $\sum_j \big( Z^\alpha \big)_{ij}T_j$ for general $\alpha$, one can use \p{Zeq} to establish relation
\be\label{ZTrels2}
\sum_j \big(Z^\alpha\big)_{ij}T_j = T_i \sum_j \big(Z^{\alpha-1}\big)_{ij}T_j-\sum_j Z_{ij}T_j \sum_k \big(Z^{\alpha-2}\big)_{jk}T_k.
\ee
Substituting here $\sum_j \big(Z^\alpha\big)_{ij}T_j = f(\alpha)\big(  T_i \big)^{\alpha+1}$, one finds that it is needed to calculate also $\sum_j Z_{ij}\big(T_j\big)^\alpha$:
\be\label{ZTrels3}
\sum_{j,k}Z_{ij}Z_{jk}\big( T_k \big)^\alpha = T_i \sum_k Z_{ik} \big( T_k \big)^\alpha - \sum_k Z_{ik} \big( T_k \big)^{\alpha+1}.
\ee
Substituting here $\sum_k Z_{ik} \big( T_k \big)^\alpha = g(\alpha)\big( T_i \big)^{\alpha+1}$, one obtains equation
\bea\label{geq}
g(\alpha) g(\alpha+1) = g(\alpha)- g(\alpha+1) \; \Rightarrow \; \left(\frac{1}{g}\right)(\alpha) = \alpha+\mbox{const} \; \mbox{or} \; g(\alpha) = \frac{1}{\alpha+1} \; \mbox{for} \; g(1)=\frac{1}{2}.
\eea
With use of \p{geq} equation \p{ZTrels2} can be reduced to single difference equation by assuming  $\sum_j \big(Z^\alpha\big)_{ij}T_j = f(\alpha)\big(  T_i \big)^{\alpha+1}$:
\be\label{feq}
(\alpha+1)f(\alpha) =(\alpha+1) f(\alpha-1) -f(\alpha-2) \; \Rightarrow \; f(\alpha) = \frac{1}{(1+\alpha)!}.
\ee
Therefore, the solution for $\lambda_i$ reads
\bea\label{lambdasol}
\lambda_i = \sum_{\alpha=0}^\infty\sum_j \big( Z^{\alpha} \big)_{ij}T_j = \sum_{\alpha=0}^\infty \frac{\big( T_i\big)^{\alpha+1}}{(1+\alpha)!} = e^{T_i} -1 \; \mbox{and}\nn \\
 \sum_j \left(\frac{1}{1-\Pi}\right)_{ij}\xi_j = e^{-\frac{\im}{2}\sum_k W(x_i-x_k)\xi_k\bxi_k}\xi_i \; \mbox{for}\; W(x) = \frac{1}{x},
\eea
which proves the formula \p{Qrat}.

\subsection*{Trigonometric and hyperbolic $W$}
In the case $c\neq 0$, the situation is more complicated. The main relation is
\be\label{ZalphaTc1}
\sum_j \big( Z^\alpha \big)_{ij}T_j = T_i \sum_k  \big( Z^{\alpha-1} \big)_{ij}T_j - \sum_k Z_{ik}T_k \sum_m \big(Z^{\alpha-2}\big)_{km}T_m + \frac{c}{4}J \sum_{k,m}\xi_k \bxi_k\, \big( Z^{\alpha-2}  \big)_{km}T_m.
\ee
At the first glance, it cannot be solved by substitution $\sum_j \big( Z^\alpha \big)_{ij}T_j =\sum_{\beta=0}^{\alpha+1} f(\alpha,\beta)\big(  T_i\big)^{\alpha+1-\beta}J^\beta$. The problem is to relate $\sum_{k,m} \xi_k\bxi_k\,  \big( Z^{\alpha-2} \big)_{km}T_m$ to other quantities.  The equation on $\sum_j Z_{ij}\big( T_j \big)^\alpha$
\be\label{ZTalphac1}
\sum_{j,k}Z_{ij}Z_{jk}\big( T_k \big)^\alpha  = T_i \sum_{j}Z_{ij}\big( T_j \big)^\alpha- \sum_{j}Z_{ij}\big( T_j \big)^{\alpha+1} + \frac{c}{4}J \sum_k \xi_k \bxi_k\, \big( T_k \big)^\alpha
\ee
is solvable by substitution
\be\label{ZTalphac1sol}
\sum_{j}Z_{ij}\big( T_j \big)^\alpha = g(\alpha)\big( T_i  \big)^{\alpha+1} + h(\alpha) \frac{c}{4}J \sum_k \xi_k \bxi_k \,\big( T_k \big)^{\alpha -1}, \;\; g(\alpha)=\frac{1}{1+\alpha}, \;\; h(\alpha)= \frac{\alpha}{1+\alpha}.
\ee
Moreover, this allows to find $\sum_k \xi_k \bxi_k \big( T_k  \big)^\alpha$:
\bea\label{xibxiTalpha}
&&\sum_k \xi_k \bxi_k \big( T_k  \big)^\alpha = -\frac{\im}{2} \sum_{k,l} \xi_k \bxi_k W(x_k-x_l)\xi_l \bxi_l \big(  T_k\big)^{\alpha-1} =   \nn \\
&&=-\sum_{k,l}\xi_l \bxi_l\, Z_{lk}\big( T_k\big)^{\alpha-1} = -\frac{1}{\alpha}\sum_l \xi_l \bxi_l\,\big(  T_l\big)^{\alpha} - \frac{\alpha -1}{\alpha}\frac{cJ}{4}\sum_l \xi_l \bxi_l\, \big(  T_l\big)^{\alpha-2}\ .
\eea
Therefore, we find
\be\label{xibxiTalpharel}
\sum_k \xi_k \bxi_k \big( T_k  \big)^\alpha = -\frac{\alpha-1}{\alpha+1} \frac{cJ^2}{4} \sum_l \xi_l \bxi_l\, \big(  T_l\big)^{\alpha-2}.
\ee
Applying \p{xibxiTalpharel} to itself, one can continue reduction
\be\label{xibxiTalpha2}
\sum_k \xi_k \bxi_k \big( T_k  \big)^\alpha =\ldots= \left( -\frac{cJ^2}{4} \right)^\beta \frac{\alpha-2\beta+1}{\alpha+1}\sum_l \xi_l \bxi_l\, \big(  T_l\big)^{\alpha-2\beta}
\ee
and reduce $\sum_k \xi_k \bxi_k \big( T_k  \big)^\alpha$ to $\sum_k \xi_k \bxi_k \big(T_k\big)^0 =J$ if $\alpha$ is even or to $\sum_k \xi_k \bxi_k \big(T_k\big)^1 =-\frac{\im}{2}\sum_{k,l}\xi_k \bxi_k W(x_k -x_l)\xi_k\bxi_l=0$ in the case of odd $\alpha$. Finally,
\bea\label{xibxiTalpha3}
\sum_k \xi_k \bxi_k \big( T_k  \big)^\alpha = \left\{  \begin{array}{l} 0,\;\; \alpha = 2\mathbb{N}+1, \\
 \frac{J}{1+\alpha} \left( - \frac{c J^2}{4} \right)^{\alpha/2}, \;\; \alpha=2\mathbb{N}. \end{array}  \right.
\eea

Using this formula, one can substitute $\sum_j \big( Z^\alpha \big)_{ij}T_j = \sum_{\beta=0}^{\alpha+1}f(\alpha,\beta)\big(  T_i\big)^{\alpha+1-\beta}J^\beta$  to \p{ZalphaTc1} and reduce  it  to a system of difference equations:
\bea\label{ZalphaTc2}
\sum_{\beta=0}^{\alpha+1} f(\alpha,\beta) \big( T_i \big)^{\alpha+1-\beta}J^{\beta} = \sum_{\beta=0}^{\alpha} f(\alpha-1,\beta) \big( T_i \big)^{\alpha+1-\beta}J^{\beta} - \nn \\
-\sum_{\beta=0}^{\alpha-1} \frac{f(\alpha-2,\beta)}{\alpha-\beta+1} \big( T_i \big)^{\alpha+1-\beta}J^{\beta} + \frac{cJ}{4}\sum_{\beta=0}^{\alpha-1}\frac{f(\alpha-2,\beta)}{\alpha-\beta+1}\sum_k \xi_k \bxi_k\, \big( T_k  \big)^{\alpha-1-\beta} J^\beta.
\eea
Counting powers of $J$ with the help of \p{xibxiTalpha3}, one can note that the last term is proportional to $J^{\alpha+1}$. This power of $J$ can be found also only in the first term. Only first and second terms contain $J^\alpha$. Thus one can write down, considering $J^\alpha$ and $J^\gamma$, $\gamma<\alpha$, two equations:
\be\label{f1f2eq}
f(\alpha,\beta) = f(\alpha-1,\beta) - \frac{f(\alpha-2,\beta)}{\alpha-\beta +1}, \;\; f(\alpha,\alpha) = f(\alpha-1,\alpha).
\ee
Solution of the first is
\be\label{f1f2sol}
f(\alpha,\beta) = \frac{a(\beta)}{(\alpha-\beta+1)!},
\ee
second is then satisfied automatically, and the only way to fix $a(\beta)$ is to consider $J^{\alpha+1}$ terms. Let us consider them separately for odd and even $\alpha$. Then, taking into account \p{xibxiTalpha3}, only even and odd $\beta$ in \p{ZalphaTc2} are relevant, respectively. Noting that $f(\alpha,\alpha+1) = a(\alpha+1)$, one can find
\bea\label{aalpha}
\alpha=2\mathbb{N}+1: && a(\alpha+1) =  \sum_{\beta=0}^{(\alpha-1)/2} \frac{(-1)^{(\alpha-1-2\beta)/2} a(2\beta)}{(\alpha-2\beta+1)!}\left(\frac{c}{4}\right)^{(\alpha+1-2\beta)/2}, \nn \\
\alpha=2\mathbb{N}: && a(\alpha+1) =  \sum_{\beta=0}^{\alpha/2} \frac{(-1)^{(\alpha-2\beta)/2} a(2\beta-1)}{(\alpha-2\beta+2)!}\left(\frac{c}{4}\right)^{(\alpha+2-2\beta)/2}.
\eea
Note that the equations relate values of $a$ of even numbers to  values of $a$ of even numbers, and the same for odd, not mixing them together. Thus, as $a(0)=1$, $a(1)=0$, one finds that $a(2\mathbb{N}+1)=0$. Relabeling $\alpha-1\rightarrow \alpha$ in the first equation, one can finally obtain the relation, that expresses $a$ of any even $\alpha$ in terms of $a(\alpha-2)$, $a(\alpha-4)$, e.t.c.
\be\label{aalphafin}
 a(\alpha+2=2\mathbb{N}) = \sum_{\beta=0}^{\alpha/2} \frac{(-1)^{(\alpha-2\beta)/2} a(2\beta)}{(\alpha-2\beta+2)!}\left(\frac{c}{4}\right)^{(\alpha+2-2\beta)/2}.
\ee
A few first of $a(\alpha)$ read
\be\label{a02468}
a(0)=1, \;  a(2) =\frac{c}{8}, \; a(4) = \frac{5c^2}{384}, \; a(6) = \frac{61c^3}{46\,080}, \; a(8) = \frac{277c^4}{2\,064\,384}, \; \ldots
\ee
Coefficients can be found this way up to any desired order. It was checked up to 20th order in $T\cdot J$ that
\bea\label{sumZalphaTc}
\lambda_i =\sum_{\alpha=0}^{\infty}\sum_{\beta=0}^{\alpha+1} f(\alpha,\beta)\big(T_i\big)^{\alpha-\beta+1}J^\beta = e^{T_i} \frac{1}{\cos\left( \frac{\sqrt{c} J}{2} \right)}-1 \; \Rightarrow \; \nn\\
 \sum_j \left(\frac{1}{1-\Pi}\right)_{ij}\xi_j =\frac{1}{\cosh\left( \frac{J}{2} \right)} e^{-\frac{\im}{2}\sum_k W(x_i-x_k)\xi_k\bxi_k}\xi_i \; \mbox{for}\; W(x) = \frac{1}{\tan x}, \nn \\
  \sum_j \left(\frac{1}{1-\Pi}\right)_{ij}\xi_j =\frac{1}{\cos\left( \frac{J}{2} \right)} e^{-\frac{\im}{2}\sum_k W(x_i-x_k)\xi_k\bxi_k}\xi_i \; \mbox{for}\; W(x) = \frac{1}{\tanh x} .
\eea
Thus the relations between the supercharges \p{Qtrig}, \p{Qhyp} hold.

\end{document}